\documentclass[prl,reprint,superscriptaddress]{revtex4-1}
\usepackage{graphicx}

\begin{document}


\title{
All-optical preparation of coherent dark states of a single rare earth ion spin in a crystal}


\author{Kangwei Xia}
\affiliation{3. Physikalisches Institut, Universit$\mathrm{\ddot{a}}$t Stuttgart, 70550 Stuttgart, Germany and \\Stuttgart Research Center of Photonic Engineering (SCoPE), 70569 Stuttgart, Germany}
\author{Roman Kolesov}
\affiliation{3. Physikalisches Institut, Universit$\mathrm{\ddot{a}}$t Stuttgart, 70550 Stuttgart, Germany and \\Stuttgart Research Center of Photonic Engineering (SCoPE), 70569 Stuttgart, Germany}
\author{Ya Wang}
\affiliation{3. Physikalisches Institut, Universit$\mathrm{\ddot{a}}$t Stuttgart, 70550 Stuttgart, Germany and \\Stuttgart Research Center of Photonic Engineering (SCoPE), 70569 Stuttgart, Germany}
\author{Petr Siyushev}
\affiliation{Institute for Quantum Optics and Center for Integrated Quantum Science and Technology (IQst), Universit$\mathrm{\ddot{a}}$t Ulm, D-89081 Germany}
\author{Rolf Reuter}
\affiliation{3. Physikalisches Institut, Universit$\mathrm{\ddot{a}}$t Stuttgart, 70550 Stuttgart, Germany and \\Stuttgart Research Center of Photonic Engineering (SCoPE), 70569 Stuttgart, Germany}
\author{Thomas Kornher}
\affiliation{3. Physikalisches Institut, Universit$\mathrm{\ddot{a}}$t Stuttgart, 70550 Stuttgart, Germany and \\Stuttgart Research Center of Photonic Engineering (SCoPE), 70569 Stuttgart, Germany}
\author{Nadezhda Kukharchyk}
\affiliation{Ruhr-Universit$\mathrm{\ddot{a}}$t Bochum, Universit$\mathrm{\ddot{a}}$tsstra$\mathrm{\beta}$e 150 Geb$\mathrm{\ddot{a}}$ude NB, D-44780 Bochum, Germany}
\author{Andreas D. Wieck}
\affiliation{Ruhr-Universit$\mathrm{\ddot{a}}$t Bochum, Universit$\mathrm{\ddot{a}}$tsstra$\mathrm{\beta}$e 150 Geb$\mathrm{\ddot{a}}$ude NB, D-44780 Bochum, Germany}
 \author{Bruno Villa}
 \affiliation{3. Physikalisches Institut, Universit$\mathrm{\ddot{a}}$t Stuttgart, 70550 Stuttgart, Germany and \\Stuttgart Research Center of Photonic Engineering (SCoPE), 70569 Stuttgart, Germany}
\author{Sen Yang}
\affiliation{3. Physikalisches Institut, Universit$\mathrm{\ddot{a}}$t Stuttgart, 70550 Stuttgart, Germany and \\Stuttgart Research Center of Photonic Engineering (SCoPE), 70569 Stuttgart, Germany}
\author{J\"{o}rg Wrachtrup}
\affiliation{3. Physikalisches Institut, Universit$\mathrm{\ddot{a}}$t Stuttgart, 70550 Stuttgart, Germany and \\Stuttgart Research Center of Photonic Engineering (SCoPE), 70569 Stuttgart, Germany}


\date{\today}

\begin{abstract}

All-optical addressing and coherent control of single solid-state based quantum bits is a key tool for fast and precise control of ground state spin qubits. So far, all-optical addressing of qubits was demonstrated only in very few systems, such as color centers and quantum dots. Here, we perform high-resolution spectroscopic of native and implanted single rare earth ions in a solid, namely a cerium ion in yttrium aluminum garnet (YAG). We find narrow and spectrally stable optical transitions between the spin sublevels of the ground and excited optical states. Utilizing those transitions we demonstrate the generation of a coherent dark state in electron spin sublevels of a single Ce$^{3+}$ ion in YAG, by coherent population trapping.

\end{abstract}


\maketitle

Coherent population trapping (CPT)~\cite{gray1978,boller1991, fleischhauer2005,harris2008} is an all-optical way of coherent manipulation of electron and nuclear spin qubits. CPT and related physical phenomena (slow light, electromagnetically induced transparency, etc.) were initially applied to quantum ensembles. In recent years, however, it has been demonstrated on single fluorescent centers in solids~\cite{santori2006, xu2008, togan2011, pingault2014, rogers2014} resulting in all-optical control of single qubits~\cite{greve2011,yale2013}, which is a significant step towards fast and high fidelity control of single qubit spins~\cite{awschalom2013,gao2015}.

Rare-earth ions residing in inorganic crystal have been widely studied and applied in fields ranging from solid-state spectroscopy and laser physics~\cite{urquhart1988} to quantum information processing~\cite{thiel2011}, due to their narrow optical transitions~\cite{sun2002, perrot2013} and long spin coherence time~\cite{longdell2005, heinze2013}. In particular, rare-earth ions in solids are promising systems for quantum information storage and processing~\cite{turukhin2001, ohlsson2002}. Achievements comprise six hours coherence time of nuclear spins in Eu:$\mathrm{Y_2SiO_5}$ crystal~\cite{zhong2015},  coherent storage of single photon states in  Nd:YVO$_4$~\cite{riedmatten2008} and entangled photon pairs in Nd:$\mathrm{Y_2SiO_5}$ crystals~\cite{clausen2011}. In addition to progress made with ensembles of rare-earth ions, the detection of individual ions have been  demonstrated in three  rare-earth  species recently~\cite{kolesov2012, kolesov2013, yin2013, utikal2014, eichhammer2015}. High fidelity spin control of single Ce$^{3+}$ ion has been demonstrated by applying a resonant microwave field~\cite{siyushev2014}. With  this  microwave control, the electron spin coherence time was extended from 150~ns to 2~ms by using the dynamical decoupling technique. Extending these finding to all-optical spin control would enable much faster control and make best use of the photon-spin coupling of rare earth ions.

\begin{figure}[h!]
\includegraphics[width=3.4in]{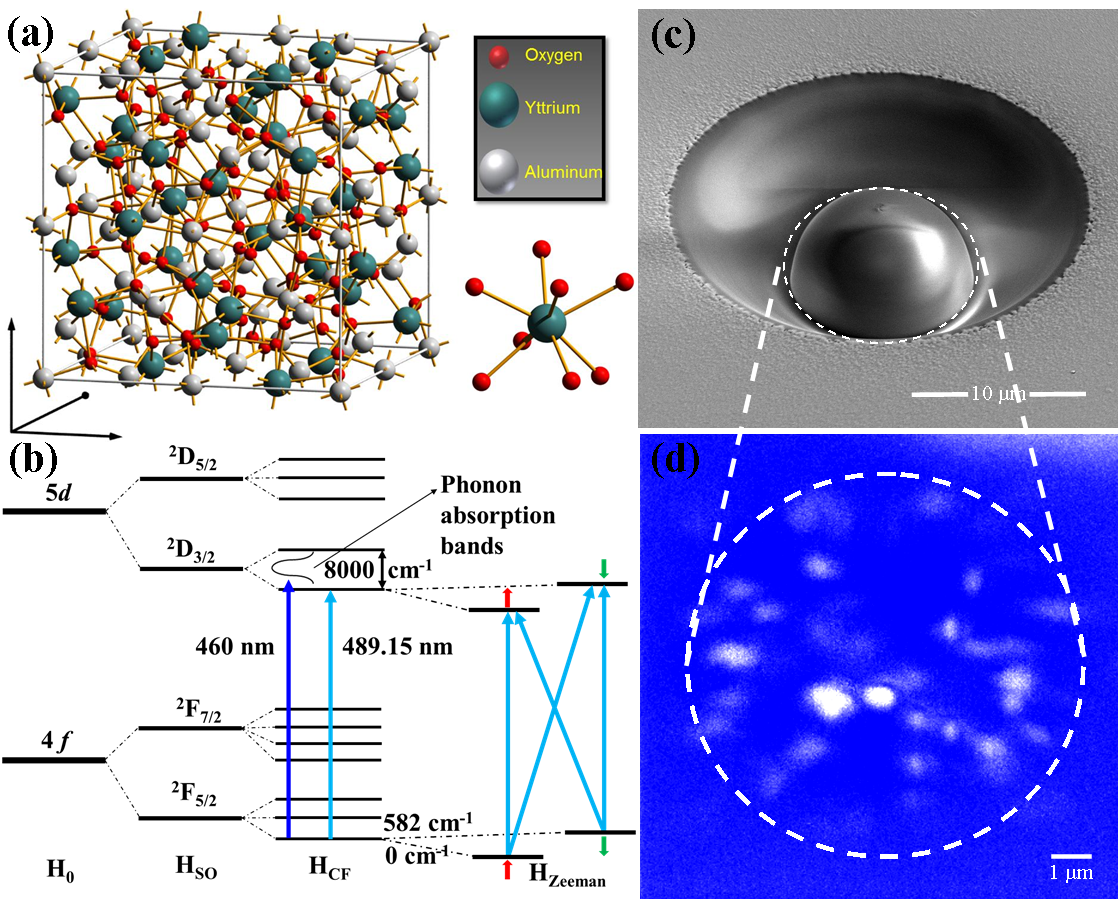}%
\caption{(a). Unit cell of a YAG crystal.  (b) Energy levels of Ce$^{3+}$ ions in YAG. (c) SEM image of a SIL on the surface of the YAG crystal. (d) Laser scanning microscopy image of Ce$^{3+}$ ions underneath the SIL. The Ce$^{3+}$ ion is excited by a 440~nm pulsed laser through the phonon absorption sideband. Bright spots correspond to individual Ce$^{3+}$ ions.\label{figure1}}
\end{figure}

\begin{figure*}
\includegraphics[width=7 in]{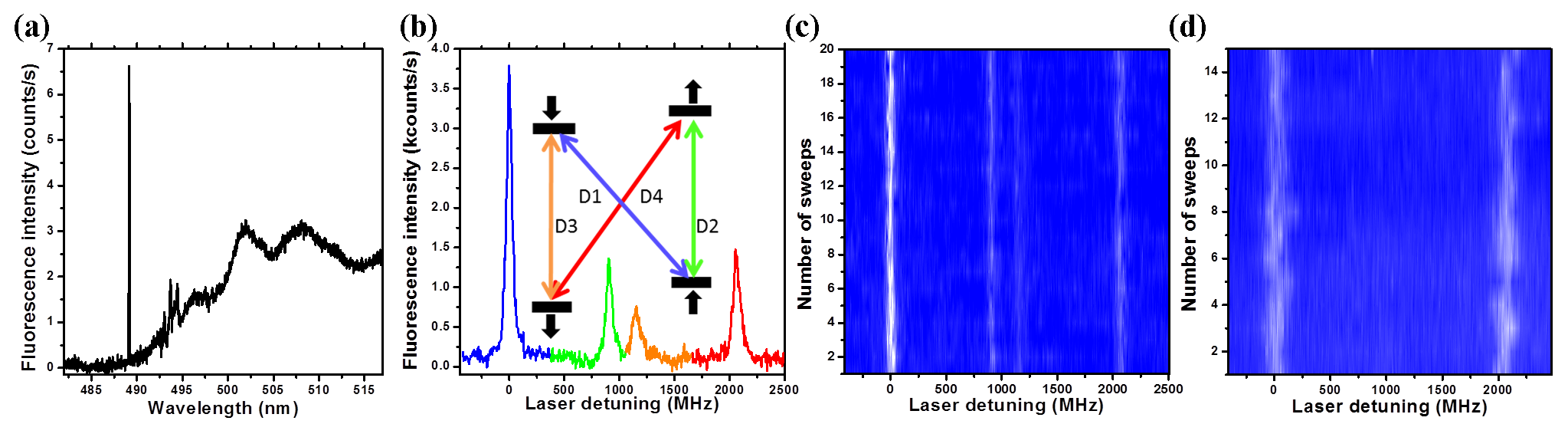}%
\caption{(a) Emission spectrum of a single Ce$^{3+}$ ion at cryogenic temperature, showing a sharp ZPL and broad phonon absorption sideband. The single Ce:YAG is excited by a 440~nm femtosecond laser with 7.6~MHz repetition rate. The spectrometer has 1800~grooves/mm grating and the integration time is 5~min. (b) Excitation spectrum of the single Ce$^{3+}$ ion. The CW diode laser with a wavelength of 489.15~nm is swept. While, the pulsed laser at 440~nm repumps the ion. An additional microwave(MW) field frequency of 1.15 GHz is applied simultaneously. The frequency matches the splitting of the lowest ground state spin transitions. (c) 20 successive photoluminescence excitation sweeps of a native single Ce ion. (d) Consecutive frequency sweeps of a single Ce ion created by focused ion implantation. \label{figure2}}
\end{figure*}

Here, we report on CPT in a "dark" coherent superposition of the electron spin sublevels of a single Ce$^{3+}$ ion in YAG based on resonant optical excitation.  CPT results in  lower fluorescence yield of a Ce$^{3+}$ center when it is excited by two laser fields  in two-photon resonance with the ground-state spin transition.
 Spectroscopic properties of resonant optical transitions of single native ions as well as single implanted ions have been studied. These studies reveal an optical linewidth of $\sim$$\mathrm{2\pi}$$\times$80~MHz and small spectral diffusion  compared to color centers in diamond~\cite{siyushev2013, chu2014} and quantum dots \cite{brunner2009}.

Figure~\ref{figure1}(a) shows the unit cell of a YAG ($\mathrm{Y_3 Al_5 O_{12}}$) crystal.
 Trivalent cerium ions substitute trivalent yttrium ions and form color centers. Cerium can be found in yttrium-containing crystals as residual impurities. Alternatively, individual Ce can be introduced into the crystal artificially by doping during the crystal growth or by ion implantation.
The energy levels of Ce$^{3+}$ in YAG are shown in Fig.~\ref{figure1}(b)~\cite{yang1996,kolesov2013}. Ce$^{3+}$ has only one unpaired electron, and its ground states are located in the 4$f^\mathrm{1}$ shell. Electrons in the 4$f$ shell are efficiently screened by closed outer lying 5$s$ and 5$p$ shells. This screening is responsible for the weak interaction between ions and their surrounding environment. The ground state is split into two sublevels $\mathrm{^2F_{5/2}}$ and  $\mathrm{^2F_{7/2}}$ due to spin-orbit coupling. These two sublevels are further split by the crystal field interaction into three and four Kramers doublets, respectively. If an external magnetic field is applied, the degeneracy of these seven Kramers doublets is lifted. The excited state is located in the 5$d$ shell. It splits into five Kramers doublets due to the combined action of the spin-orbit coupling and the crystal field. The energy difference between the two lowest 5d Kramers doublets is approximately 8,000~cm$^{-1}$~\cite{yang1996,kolesov2013} and, therefore, the excited states can be optically addressed individually. In addition, the quantum efficiency of the 5$d \rightarrow$ 4$f$ transitions is close to 100\%~\cite{weber1973}. The lifetime of the lowest 5$d$ state is 60~ns~\cite{hamilton1989,kolesov2013}.

In the experiment, a [1 1 0] orientated ultrapure YAG  crystal (Scientific Materials) is used. An external magnetic field ($B\approx 450$~G)  is applied perpendicular to the laser beam direction, so that four optical transitions between the two pairs of spin states of the 4$f$ and 5$d$ levels are allowed, as shown in Fig.~\ref{figure1}~(c)~\cite{kolesov2013}. A $\Lambda$ scheme, which is a requirement for CPT, can be formed by optically mixing both ground states to either of the excited spin states.
Single Ce$^{3+}$ ions are detected under a home-built high resolution confocal microscope at cryogenic temperature ($T \approx 3.5$~K).
A solid immersion lens (SIL) is milled by focused ion beam on the surface of the YAG crystal to improve the spatial resolution and the fluorescence collection efficiency of the confocal microscope~\cite{hadden2010}(see Fig.~\ref{figure1}(c)). The laser scanning microscopy image shown in Fig.~\ref{figure1}(d) is obtained through non-resonant pulsed excitation (440 nm) with a frequency-doubled femtosecond Ti:Sapphire laser. Each bright spot corresponds to  an individual native Ce$^{3+}$ ion. The emitted photons associated with broad phonon sideband are detected by an avalanche photodiode (APD) in a spectral range between 500-625 nm (see our previous work~\cite{siyushev2014}). A tunable single-mode narrow linewidth ($\sim$500~kHz) continuous wave (CW) laser (wavelength of 489.15 nm, Toptica Photonics DL pro) is used to resonantly excite single Ce ions.

\begin{figure}[hhh]
\includegraphics[width=3.4in]{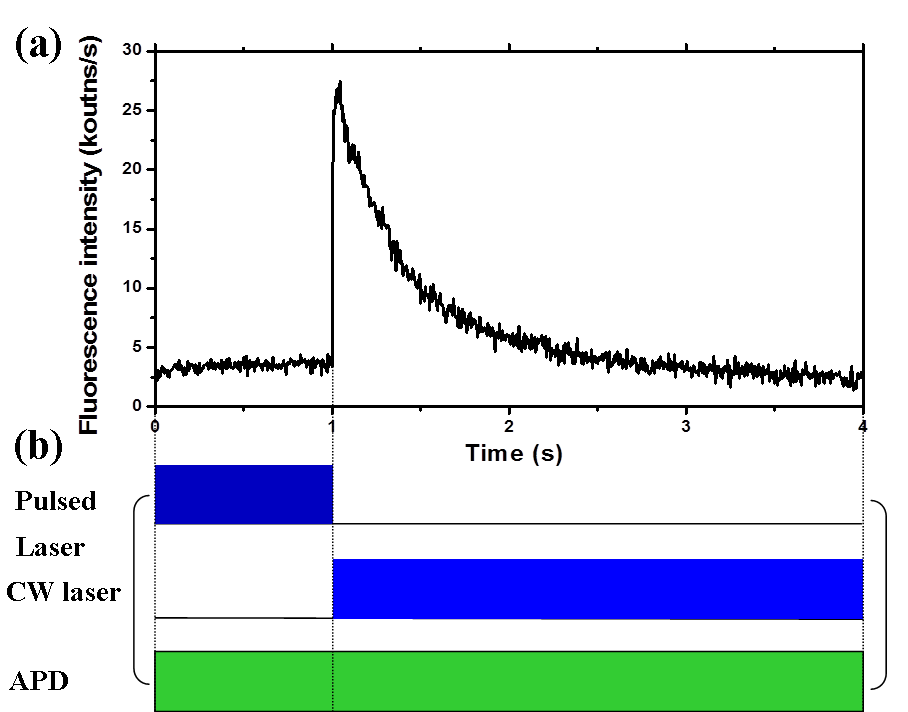}%
\caption{(a) Fluorescence intensity of a single Ce ion under pulsed and CW laser excitation. (b) Scheme of the laser pulse sequences. The pulsed laser is used to bring the Ce$^{3+}$ back and the Ce:YAG is ionized by CW laser excitation when the pulsed laser is switched off. The femtosecond pulsed laser wavelength is 448~nm. The repetition rate is 2.5~MHz and the average laser power is $\mathrm{10~\mu W/cm^2}$. A blue diode laser (451~nm) with $\mathrm{150~\mu W}$ power is used as the CW laser. \label{figure3}}
\end{figure}

A portion of the emission spectrum of a single Ce$^{3+}$ ion in the vicinity of its zero-phonon line (ZPL),  measured by a high resolution spectrometer, is shown in Fig.~\ref{figure2}(a). A sharp zero-phonon line (ZPL) is located at 489.15~nm and accompanied by a red-shifted phonon sideband partly shown in the figure~\cite{bachmann2009}.

A CW laser  is swept across the ZPL position to obtain the excitation spectrum of the single Ce ion shown in Fig.~\ref{figure2}(b). Four individual optical excitation-transition lines are well resolved. These lines correspond to four different optical transitions between the lowest Kramers doublets of the ground state and of the excited state with the assignment indicated in Fig.~\ref{figure2}(b).  The full width at half maximum(FWHM) of the optical transitions is $\sim$$\mathrm{2\pi}$$\times$80~MHz, which is broader than the lifetime limited linewidth $\mathrm{2\pi}$$\times$4~MHz. This broadening is caused mainly  by the strong $^{27}$Al nuclear spin bath, i.e. it represents an intrinsic property of the host material~\cite{siyushev2014}.

By monitoring  the fluorescence during each successive frequency sweep through the resonant transitions,  we observe stable optical resonance lines without obvious spectral diffusion, as shown in Fig~\ref{figure2}(c). It indicates that native single  Ce ions have a surprisingly good spectral stability under resonant excitation.

In addition to the native single Ce centers, the spectral stability of Ce ions created by ion implantation has been investigated (see Fig.~\ref{figure2}(d)). With high-dose implantation, we found about 100 ions in one confocal spot(see Supplemental Material).
The optical transition linewidth of the implanted Ce ions is $\sim$$\mathrm{2\pi}$$\times$150~MHz, increased mainly due to extra strain introduced by ion implantation.
Since this linewidth is much narrower than the inhomogeneous width of $\sim$$\mathrm{2\pi}$$\times$550~GHz, it is possible to address single Ce$^{3+}$ ions by tuning the excitation laser wavelength into resonance with the optical transitions. Compared to other solid-state systems, e.g. defects in diamond and quantum dots~\cite{pingault2014, rogers2014, siyushev2013, chu2014,xu2008}, implanted single Ce$^{3+}$ presents narrow optical transitions and a stable spectrum. The combination of narrow optical transitions and spectral stability makes precision optical control of single Ce ion spins possible comparable to single praseodymium ions in solids~\cite{utikal2014, eichhammer2015, nakamura2014}.

In order to measure the  excitation spectrum of a single Ce$^{3+}$,  a  CW laser is applied to resonant excitation as well as a low repetition rate,  femtosecond laser with high peak intensity. If a single Ce$^{3+}$ is excited  with CW laser only, its fluorescence intensity shows  a smooth decay and quickly goes to background level in a few seconds, as shown in Fig.~\ref{figure3}(a). The decay curve in Fig.~\ref{figure3}(a) is the observation of a fluorescence time traces of the single Ce ion under CW laser excitation only. Surprisingly, in contrast to all other single emitters a gradual decay of the fluorescence is observed and not as usual a step-wise bleaching.
In order to explain such gradual bleaching of a single Ce$^{3+}$ ion, we propose a model involving two competing processes: 1) photo-ionization from Ce$^{3+}$ into Ce$^{4+}$ and 2) restoration of Ce$^{3+}$ by taking an electron from a nearby deep donor. As long as there are enough deep donors in the vicinity of the ion, the cerium remains in its trivalent state and fluoresces. However, a gradual reduction of the number of donors reduces the probability of restoring the trivalent state of cerium. This leads to a gradual decrease of the fluorescence intensity.
The charge dynamics of single ions observed here is consistent with previous observations in ensembles~\cite{hamilton1989,miniscalco1978,pavlov2013}.
It also explains why attempts of detecting single Ce ions under CW laser excitation were unsuccessful~\cite{kolesov2013}.

Surprisingly,  a femtosecond laser featuring a high peak intensity restores the population of donors, which helps the Ce ion pumping back to the Ce$^{3+}$ charge state  (see Fig.~\ref{figure3}(a) and (b)).
Therefore, to keep the Ce ion photostable, we apply CW and pulsed lasers simultaneously in the experiments.
Details of these charge dynamics are discussed in the Supplementary Material.

\begin{figure}
\includegraphics[width=3.4in]{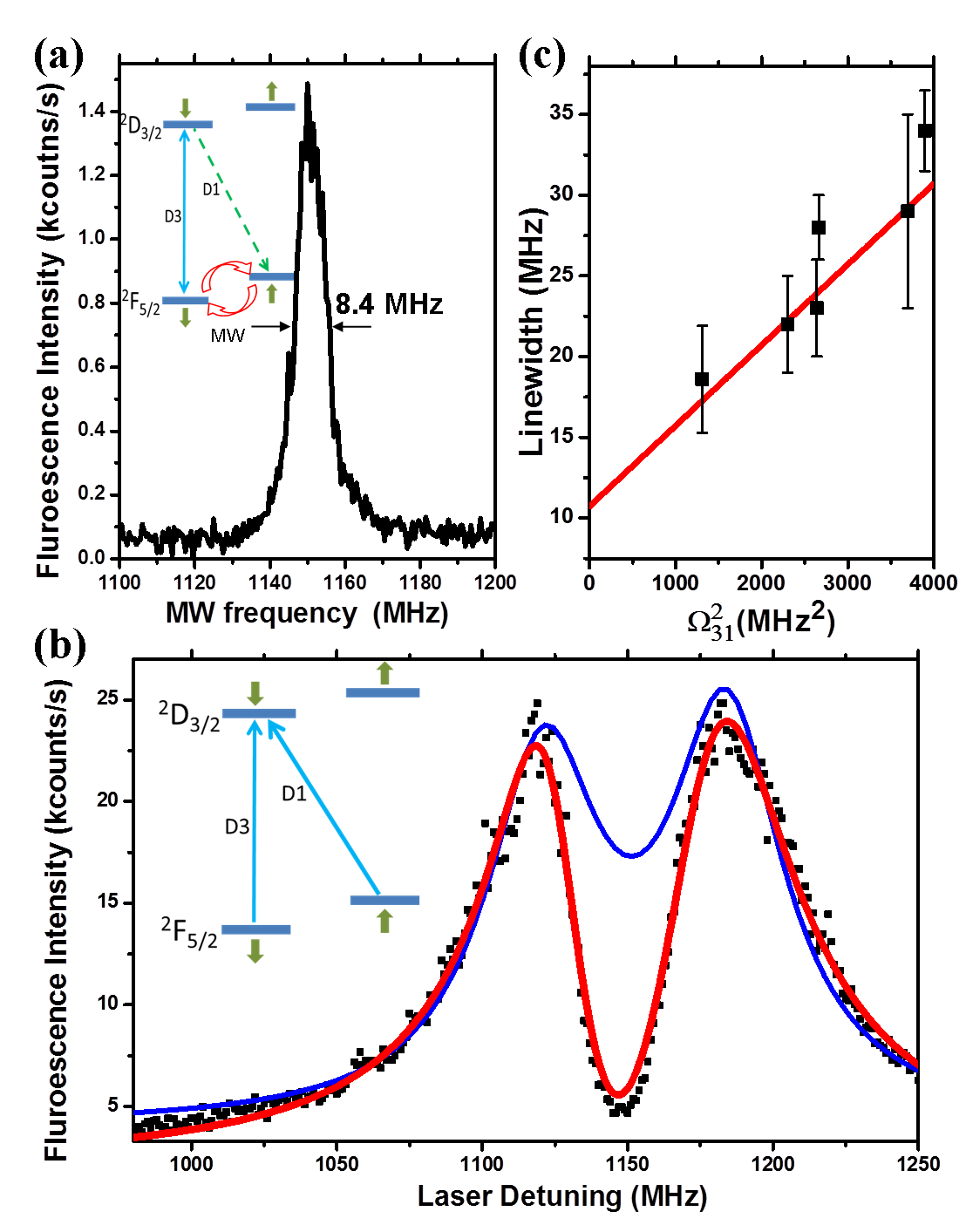}%
\caption{(a)  ODMR of the single Ce$^{3+}$ ion. During the MW frequency sweep, the CW laser is turned on to resonantly excite the Ce:YAG through the D3 transition.  Insert figure: scheme of a single Ce$^{3+}$ ion ODMR. (b) CPT of the single Ce$^{3+}$ ion. The frequency of the pump laser is the same as the one of the D1 transition. The probe laser is passed through three acousto-optic modulators to have 980~MHz to 1250~MHz tuning range to sweep around the D3 transition. The pulsed laser is electrically chopped with 2.5~kHz repetition rate  to avoid ionizing the Ce ion to the Ce$^{4+}$ state. The red curve corresponds to the fitting of this CPT process. The blue curve shows the result of a Rabi splitting simulation. (c) The linewidth of CPT dips as a function of pump laser induced Rabi frequency. \label{figure4}}
\end{figure}

From the four different optical transitions of single Ce ions, a $\Lambda$ system can be formed, consisting of two ground states and either one of the excited states. In experiments, we choose the $\Lambda$ system with transitions D1 and D3. To observe CPT, the pump laser frequency is fixed on the transition line D1, and the frequency of the probe laser is swept around the frequency of the D3 transition.
Simultaneously, the fluorescence intensity of the single Ce$^{3+}$ ion is monitored, which is shown in Fig.~\ref{figure4}(b). It contains a broad peak with a  dip going down nearly to the background level. The total width of the peak is consistent with the optical transition linewidth (Fig.~\ref{figure2}(b)). The dip is centered exactly at the D3 transition, indicating that the ground state population is coherently trapped in a dark state.

The observed dip width is around $\mathrm{2\pi}$$\times$35~MHz, caused by several sources of decoherence, including the intrinsic linewidth of the ground state spin transition and the laser power induced broadening. To understand this $\mathrm{2\pi}$$\times$35~MHz CPT linewidth, we perform optically detected magnetic resonance (ODMR) on the ground states, to obtain the intrinsic linewidth of the ground state spin transition. We use laser excitation resonant with the D3 transition to initialize the ion into the spin up state.
Microwave (MW) radiation is applied through the wire located next to the position of the ion under investigated. Then, the MW frequency is swept through resonance of the ground-state spin transition.
The power of both, laser and MW is kept low to avoid power broadening (tens of microwatts of laser power and $\sim$1~dBm microwave power according to the waveguide structure).
For the off-resonant MWs, the electron spin stays in the spin up state and the fluorescence level is low. Once the MW field is in resonance with the spin transition, the electron is  pumped back to the spin down state, resulting in a higher fluorescence yield (see Fig.~\ref{figure4}(a)).  From the observed high-contrast ODMR signal, we deduce a fidelity of the initialization of  more than 98\%. The initialization fidelity is much higher than the 50.2\% population at 3.5~K.
The inhomogeneous linewidth of the spin transition is $\mathrm{2\pi}$$\times$8.4~MHz in agreement with our previous measurements \cite{siyushev2014}. This value is smaller than the linewidth of the CPT dip, indicating the resonance is broadened by laser power.


By  fitting the measured fluorescence intensity with the solution for the density matrix equations of the four-level system, we obtain excellent agreement between theory and experimental results (see Supplementary Material). From the fits, we can estimate the driving strength to be $\mathrm{2\pi}$$\times$62~MHz for the D1 transition. The pump laser induced Rabi splitting with a much weaker dip, as shown in Fig.~\ref{figure4}(b), further indicates that the observed dip corresponds to the successful formation of a coherent dark state.

To quantify the power broadening effect, we measured the CPT dip width for various laser powers. The linewidths  are linearly dependent on the laser power, as shown in Fig.~\ref{figure4}(c), in agreement with expectations. Through a linear fitting, we extract a linewidth $\sim$10.7$\pm$3.2~MHz without laser broadening. The value is consistent with an intrinsic linewidth 8.4~MHz, obtained from the ODMR measurement.

In conclusion, we showed  narrow-linewidth, resonant transitions and good spectral stability for both native and focused ion beam implanted single Ce ions. Based on these optical properties, we demonstrated all optical formation of steady-state coherent dark states of single Ce$^{3+}$ by the CPT technique.
All-optical control of a single spin qubit based on a Ce$^{3+}$ can thus be realized by dynamically manipulating coherent laser fields.
In addition, on-chip photonic circuits for this system which adds another critical element for their use in quantum technology.

\begin{acknowledgments}
We would like to thank Philippe Goldner, Alban Ferrier, Rainer St\"{o}hr and Nan Zhao for discussions. The work is financially supported by ERC SQUTEC, EU-SIQS SFB TR21 and DFG KO4999/1-1
\end{acknowledgments}

\end{document}